\documentclass[aps,prl,twocolumn]{revtex4}
\usepackage{graphicx}

\newcommand{\grl}{\langle\!\langle}
\newcommand{\grr}{\rangle\!\rangle}

\begin{document}
\title{Microscopic theory of Gilbert damping in metallic ferromagnets}
\author{A. T. Costa}
\affiliation{Instituto de F\'isica, Universidade Federal Fluminense, 
24210-346 Niter\'oi, RJ, Brazil.}
\author{R. B. Muniz}
\affiliation{Instituto de F\'isica, Universidade Federal Fluminense, 
24210-346 Niter\'oi, RJ, Brazil.}

\begin{abstract}
We present a microscopic theory for magnetization relaxation
in metallic ferromagnets of nanoscopic dimensions that is
based on the dynamic spin response matrix in the presence
of spin-orbit coupling. Our approach allows the calculation
of the spin excitation damping rate even for perfectly crystalline
systems, where existing microscopic approaches 
fail. We demonstrate that the relaxation properties are \textbf{not} completely
determined by the transverse susceptibility alone, and that
the damping rate has a non-negligible frequency dependence
in experimentally relevant situations. Our results indicate that 
the standard Landau-Lifshitz-Gilbert phenomenology is not always appropriate to describe 
spin dynamics of metallic nanostructure  in the presence of strong spin-orbit coupling.
\end{abstract}

\maketitle


Magnetization relaxation in metals is at the heart of spin current generation and 
detection processes currently under investigation, many of them candidates to 
play protagonist roles in innovative spintronic devices. The Landau-Lifshitz-Gilbert (LLG) 
equation is widely used to describe the spin dynamic properties of magnetic materials~\cite{gilbert_ieee,Bauer_RevModPhys}. 
It includes an important system-dependent parameter, called the Gilbert damping constant, usually denoted by $\alpha_G$, 
that regulates the relaxation of the magnetization towards stability, after it is driven out of equilibrium. 
Recently, a lot of effort has been put into the determination of this damping rate~\cite{Bauer_RevModPhys,Kambersky2007,GarateMacdonald_2009I,GarateMacdonald_2009II,Bauer_Tserkovnyak_Kelly_2010,Kelly_2011,Edwards_2014}, which characterizes the pumping and absorption of pure spin currents in nanostructures that are of great interest in the field of spintronic. 
In most of them spin-orbit interaction is significant, and responsible for a desirable interplay between charge spin and angular momentum excitations. 

There is a general agreement between practitioners in the field
that  a proper microscopic theory of magnetization 
relaxation in metals requires a good description of the electronic 
structure of the system \textbf{including} spin-orbit coupling~\cite{Kambersky2007,GarateMacdonald_2009I,GarateMacdonald_2009II,Bauer_Tserkovnyak_Kelly_2010,Kelly_2011,Edwards_2014}.
The conventional approach is to combine a realistic electronic structure with some kind of
adiabatic approximation to derive expressions that can be directly related to the 
Landau-Lifshitz-Gilbert phenomenology. This strategy has been employed by  
Kambersk\'y~\cite{Kambersky2007} and many others since
\cite{GarateMacdonald_2009I,GarateMacdonald_2009II,Bauer_Tserkovnyak_Kelly_2010,Kelly_2011,Edwards_2014}.
This conventional approach has important limitations. It neglects the coupling between
transverse spin, longitudinal spin and charge excitations 
(which is an important consequence of the spin-orbit coupling), and incorrectly
pedicts the divergence of the damping parameter for a perfectly 
crystalline system.
Actually, for ferromagnets that display rotation symmetry in spin space, 
the Goldstone theorem ensures that any experiment which measures the total 
transverse magnetic moment of the sample will produce a resonant response 
with zero linewidth~\cite{MunizMills2003}. In the presence of spin-orbit 
interaction, however, this symmetry is explicitly broken, and the resonant 
spectrum acquires a finite linewidth~\cite{Costa_SOC_2010}.

We put forward a more fundamental microscopic approach to the calculation of the 
spin dynamics damping rate that takes fully into account the effects of SOC 
on the spectrum of spin excitations of itinerant systems. Namely, we consider
the coupling of transverse spin excitations to longitudinal spin and charge excitations,
induced by the spin-orbit interaction. We calculate the FMR spectrum at finite frequencies 
and arbitrary anisotropy values, without employing any adiabatic approximation. 
We will show that those ingredients are essential to correctly describe the magnetization 
relaxation in very clean metallic ferromagnets of
nanoscopic dimensions, and that the Landau-Lifshitz-Gilbert
phenomenology fails to capture essential features of the magnetization 
dynamics in those systems. 

This letter is organized as follows: we will present briefly our formalism, discuss its main features 
and present numerical results for two model systems that illustrate common
but qualitatively different situations.


\textit{General Formalism - }The spectrum of spin excitations of a ferromagnet can be obtained from the 
spectral density associated with the transverse spin susceptibility
\begin{equation}
\chi^{+-}(l,l';\Omega) = 
\int dt e^{i\Omega t}\grl S^+_l(t),S^-_{l'}(0)\grr,
\end{equation}
where
\begin{equation}
\grl S^+_l(t),S^-_{l'}(0)\grr\equiv
-i\theta(t)\langle[S^+_l(t),S^-_{l'}(0)]\rangle,
\end{equation}
and
\begin{equation}
S^+_l=
\sum_{\mu}a^\dagger_{l\mu\uparrow}a_{l\mu\downarrow}.
\end{equation}
The operator $a^\dagger_{l\mu\sigma}$ creates one electron in the 
atomic basis state $\mu$ localized at lattice site $l$ 
with spin $\sigma$. Although we are usually interested in
$\chi^{+-}(l,l';\Omega)$ as defined above, its 
equation of motion involves the orbital-resolved susceptibility,
\begin{equation}
\chi^{+-}_{\mu\nu\mu'\nu'}(l,l';t)\equiv
\grl a^\dagger_{l\mu\uparrow}(t)a_{l\nu\downarrow}(t),
a^\dagger_{l'\mu'\downarrow}(0)a_{l'\nu'\uparrow}(0)\grr .
\end{equation}
In the absence of spin-orbit coupling (SOC) and within the random phase approximation (RPA),
the equation of motion for $\chi^{+-}_{\mu\nu\mu'\nu'}(l,l';t)$ is closed and
$\chi^{+-}(l,l';\Omega)$ can be expressed in the well-known RPA form,
\begin{equation}
\chi^{+-}(\Omega) = [1+U\chi^{+-}_0(\Omega)]^{-1}\chi^{+-}_0(\Omega)
\label{RPA_noSOC}
\end{equation}
where $\chi^{+-}_0(\Omega)$ is the mean-field (sometimes called non-interacting, or Hartree-Fock)
susceptibility. This expression is schematic and must be understood as a matrix in orbital and
site indices, in real space, or a wave-vector dependent matrix in reciprocal 
space. The crucial point, 
however, is that, in the absence of spin-orbit coupling, within the RPA, the transverse spin susceptibility is uncoupled from any other 
susceptibility. This ceases to be true when SOC is included, 
as we demonstrated in ref.~\onlinecite{Costa_SOC_2010}: $\chi^{+-}$ becomes coupled to three other 
susceptibilities, namely
\begin{eqnarray}
\chi^{(2)}_{\mu\nu\mu'\nu'}(l,l';t)\equiv
\grl a^\dagger_{l\mu\uparrow}(t)a_{l\nu\uparrow}(t),
a^\dagger_{l'\mu'\uparrow}(0)a_{l'\nu'\uparrow}(0)\grr,\\
\chi^{(3)}_{\mu\nu\mu'\nu'}(l,l';t)\equiv
\grl a^\dagger_{l\mu\downarrow}(t)a_{l\nu\downarrow}(t),
a^\dagger_{l'\mu'\downarrow}(0)a_{l'\nu'\downarrow}(0)\grr ,\\
\chi^{(4)}_{\mu\nu\mu'\nu'}(l,l';t)\equiv
\grl a^\dagger_{l\mu\downarrow}(t)a_{l\nu\uparrow}(t),
a^\dagger_{l'\mu'\uparrow}(0)a_{l'\nu'\downarrow}(0)\grr .
\end{eqnarray}
The system of equations of motion obeyed by these four susceptibilities can be cast into
a form strongly resembling the RPA result by introducing a block-vector
$\vec{\chi}\equiv (\chi^{(1)},\chi^{(2)},\chi^{(3)},\chi^{(4)} )^T$,
with $\chi^{(1)}\equiv\chi^{+-}$. With an equivalent definition for the mean-field 
susceptibilities $\chi^{(m)}_0$ we write
\begin{equation}
\vec{\chi}(\Omega) = \vec{\chi}_0(\Omega) - \Lambda \vec{\chi}(\Omega),
\end{equation}
where the ``super-matrix'' $\Lambda$ is proportional to the effective 
Coulomb interaction strength and involves convolutions of single 
particle Green functions. Explicit forms for its matrix elements
are found in Ref.~\onlinecite{Costa_SOC_2010}. The numerical
analysis of the susceptibilities $\chi^{(2)}$, $\chi^{(3)}$ and $\chi^{(4)}$ show 
that their absolute values are many orders of magnitude
smaller than those of $\chi^{(1)}=\chi^{+-}$. 
It is, thus, tempting to argue that the transverse susceptibility
is approximately decoupled from $\chi^{(2)}$, $\chi^{(3)}$ and $\chi^{(4)}$ 
and that it can be calculated
via the usual RPA expression with the single particle Green functions obtained with
spin-orbit coupling taken into account. This is not a good 
approximation in general, since the matrix elements of $\Lambda$ that couple 
$\chi^{(1)}$ to the other susceptibilities are far from negligible. 
Our numerical calculations indicate that they are essential to determine correctly the 
features of the FMR mode around the resonance frequency. Thus, the 
behaviour of $\chi^{(1)}$ in the presence of spin-orbit coupling cannot 
be inferred from $\chi_0^{(1)}$ in the zero-frequency limit
alone, as it is usually assumed in the literature on the calculation of the Gilbert damping parameter~\cite{Kambersky2007,GarateMacdonald_2009I,GarateMacdonald_2009II,Edwards_2014,Kelly_PRL2014}.

\begin{figure}
\includegraphics[width=\columnwidth]{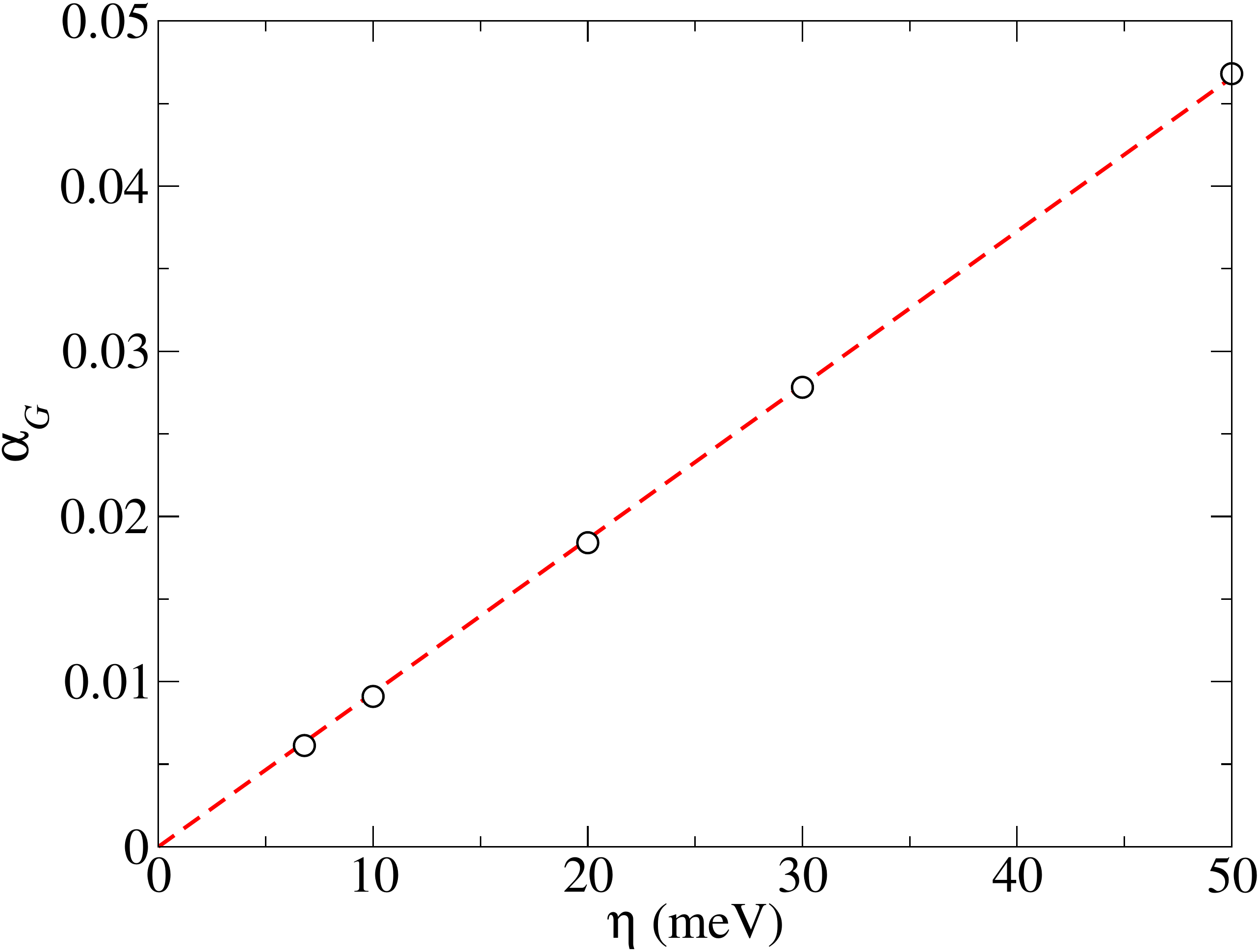}
\caption{Gilbert damping constant $\alpha_G$ as a function of the imaginary part $\eta$ added to the
real energy, for an ultrathin film of two atomic layers of Co where SOC has been turned off. 
It is clear that $\alpha_G$ vanishes as $\eta\rightarrow0$.}
\label{alpha_Co_no_SOC}
\end{figure}

\textit{Numerical Results - } We start the discussion by presenting results for the Gilbert 
constant $\alpha_G$ for unsupported ultrathin Co films. Here we determine
$\alpha_G$ from the ratio between the FMR linewidth $\Delta\Omega$ and
the resonance frequency $\Omega_0$. First we 
turn off spin-orbit coupling to check the consistency of our approach. Even
with SOC turned off we still find a finite linewidth for the FMR mode. 
It comes, as we will shortly demonstrate, from the small imaginary part 
$\eta$ that is usually added to the energy in the numerical calculations
of the single particle Green functions, in order to move their poles from
the real axis. We calculate $\alpha_G$ for various 
values of $\eta$ and extrapolate to $\eta\rightarrow0^+$, as shown in 
Fig.~\ref{alpha_Co_no_SOC}. It is clear that  $\lim_{\eta\rightarrow0^+}\alpha_G=0$.  
Thus, our approach correctly predicts that the Gilbert damping 
constant vanishes in the absence of SOC, as it should. Indeed, 
it is easy to show~\cite{MunizMills2003} that the FMR mode is a stationary 
state of the mean-field hamiltonian and, as such,
has infinite lifetime in the limit $\eta\rightarrow 0^+$. 
Now we discuss the dependence of $\alpha_G$ on $\eta$ for a fixed, 
non-zero value of the spin-orbit coupling strength $\xi$. We used 
LCAO parameters appropriate for bulk Co to describe the electronic structure
of all Co films we investigated. The quantitative details 
of the ferromagnetic ground state and excitation spectra are sensitive 
to the LCAO parameters used, but their qualitative behaviour
is very robust to small changes in the electronic structure. Our strategy is to use
the same set of LCAO parameters for all film thicknesses to 
avoid modifications in $\alpha_G$ coming directly from changes in the LCAO 
parameters. This allows us to focus on geometric effects and
on the $\eta$-dependence.

\begin{figure}
\includegraphics[width=0.9\columnwidth]{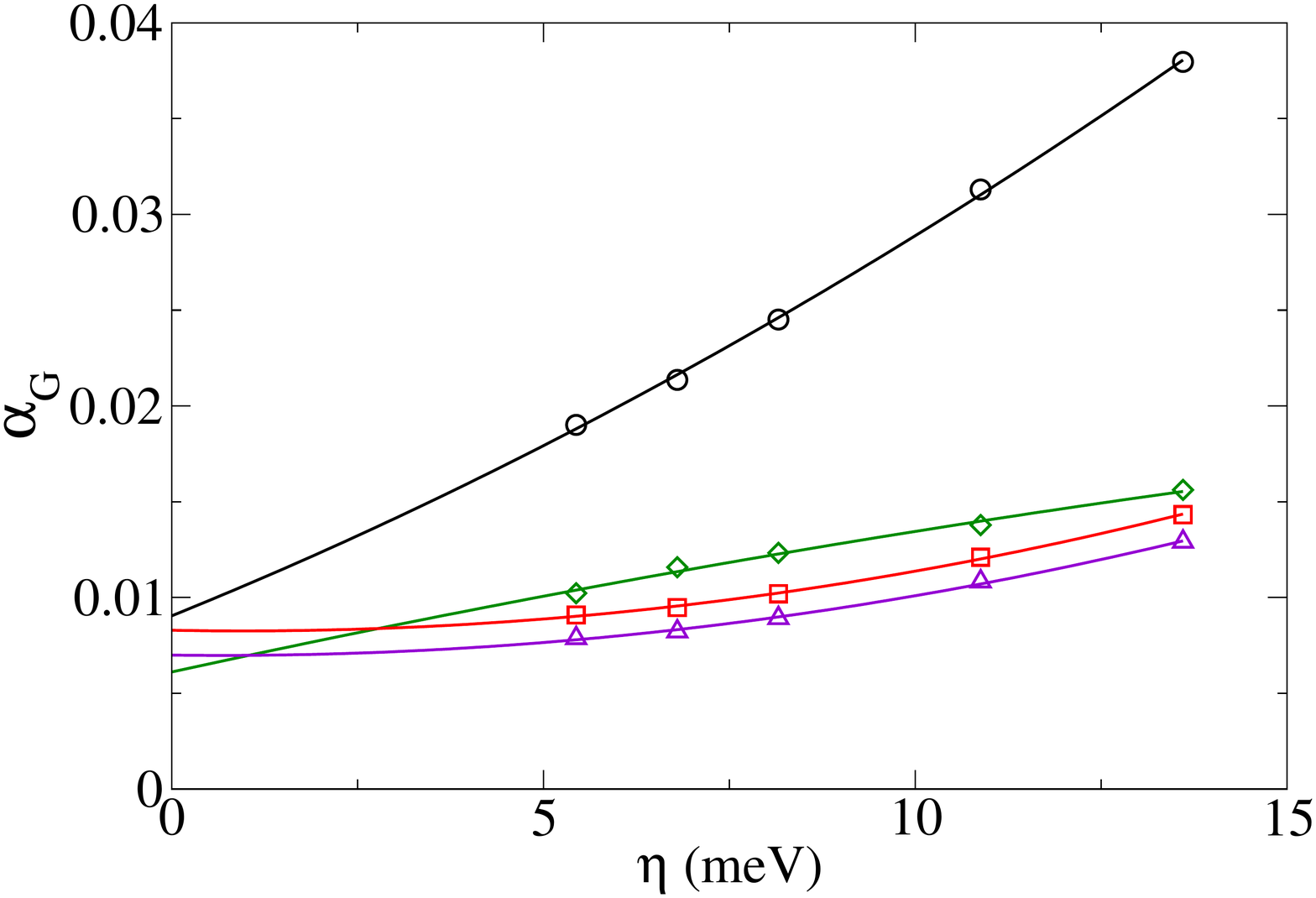}
\caption{Gilbert damping constant $\alpha_G$ as a function of the 
imaginary part $\eta$ added to the energy, for Co ultra thin films 
of various thicknesses: 1 (circles), 2 (squares), 4 (diamonds) 
and 6 (triangles) atomic layers. 
The strength of the SOC is $\xi=85$~meV. The solid lines are guides to the eye.}
\label{alpha_x_eta_x_NCo}
\end{figure}

\begin{figure}
\includegraphics[width=0.9\columnwidth]{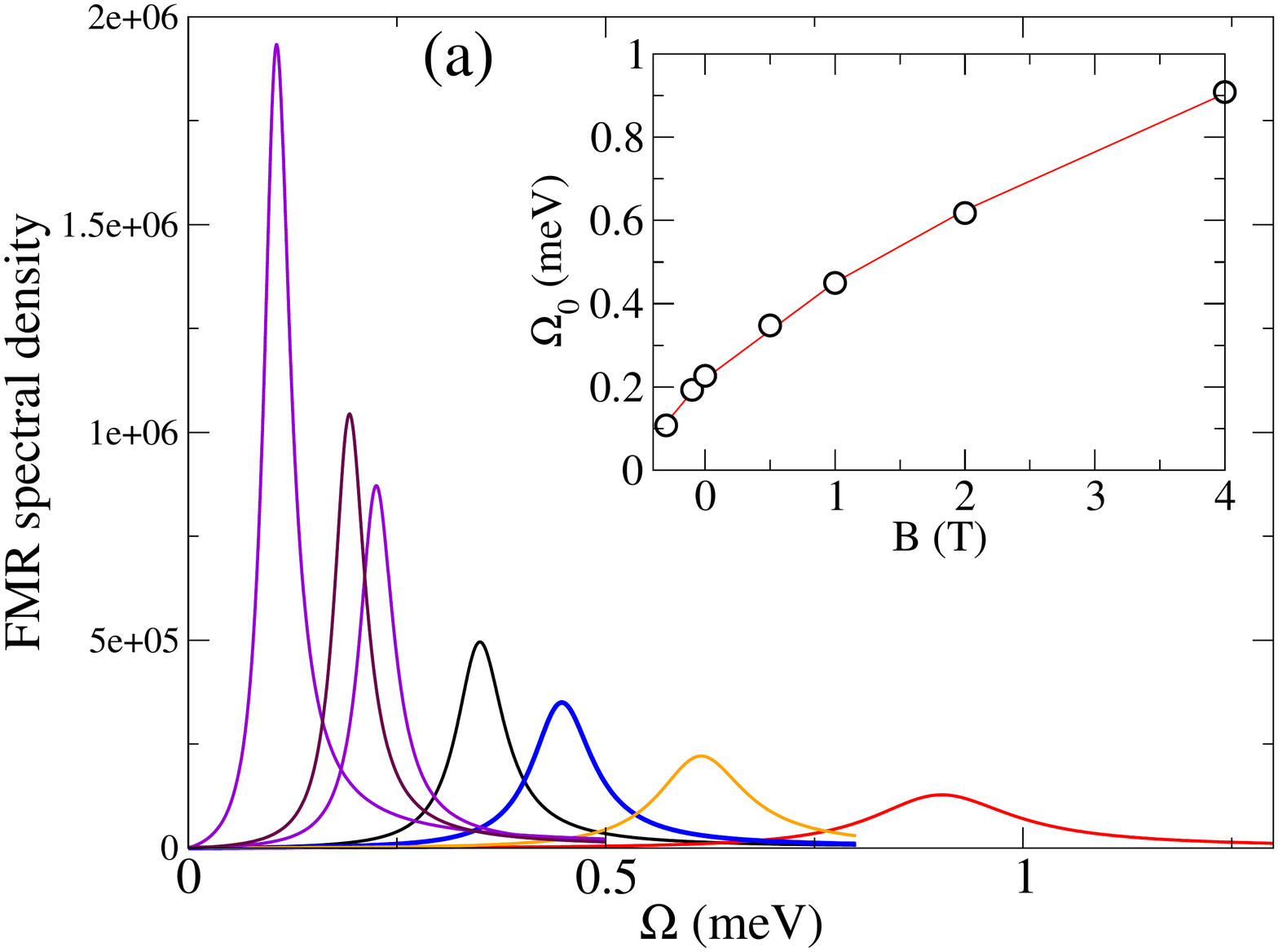}
\includegraphics[width=0.9\columnwidth]{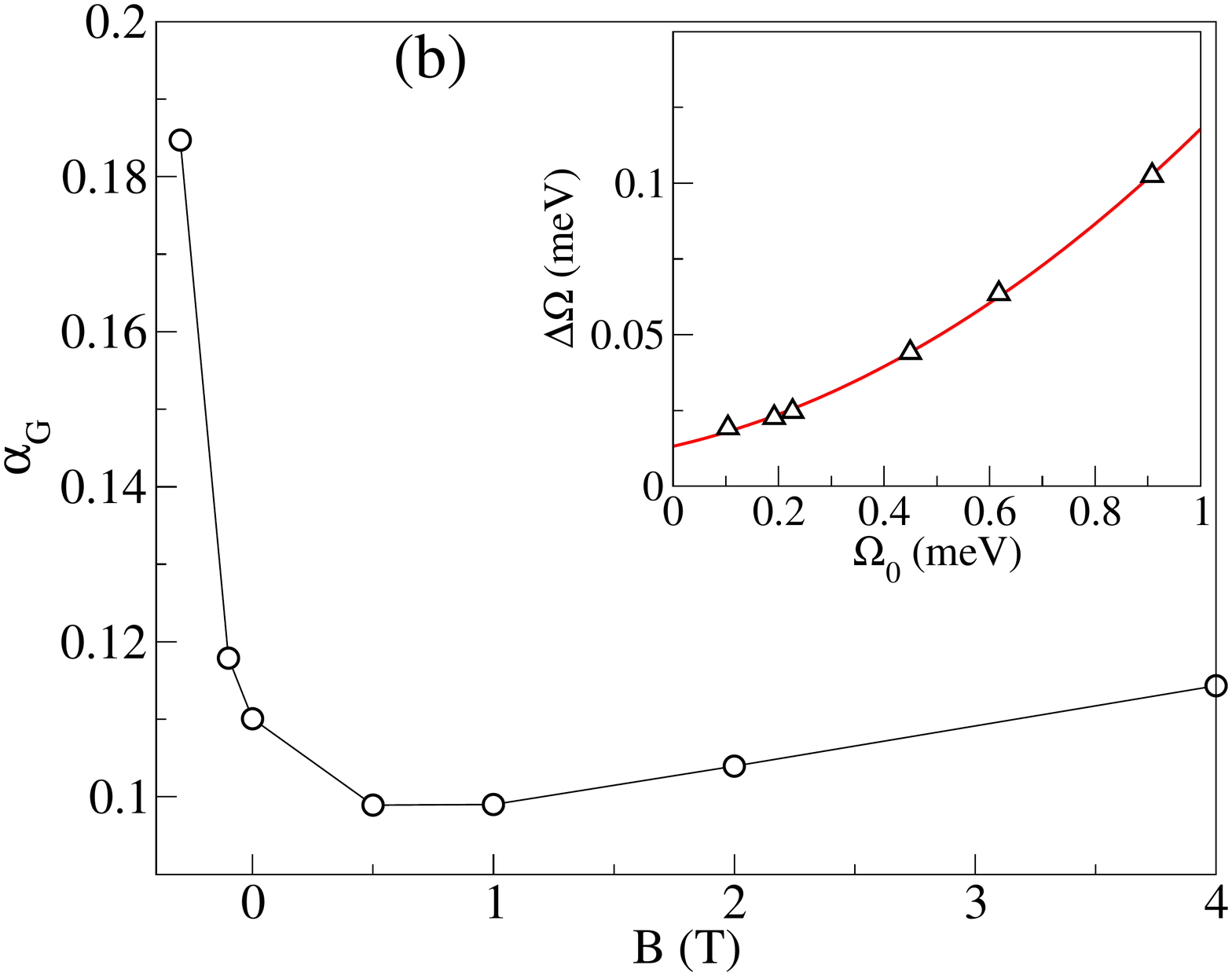}
\caption{a) Spectral densities of the FMR mode for the 2Co/2Pt system
subjected to various static magnetic fields (from -0.3 T to 4 T). The inset shows the 
resonance frequency as a function of the Zeeman field $B$.
b) The Gilbert damping parameter $\alpha_G$ as a function of
applied Zeeman field B. The inset shows the FMR line width as a function
of resonance frequency $\Omega_0$.
The strengths of the SOC are $\xi_\mathrm{Co}=85$~meV and 
$\xi_\mathrm{Pt}=600$ meV.} 
\label{DW_x_W0_2Co2Pt}
\end{figure}

\begin{figure}
\includegraphics[width=0.9\columnwidth]{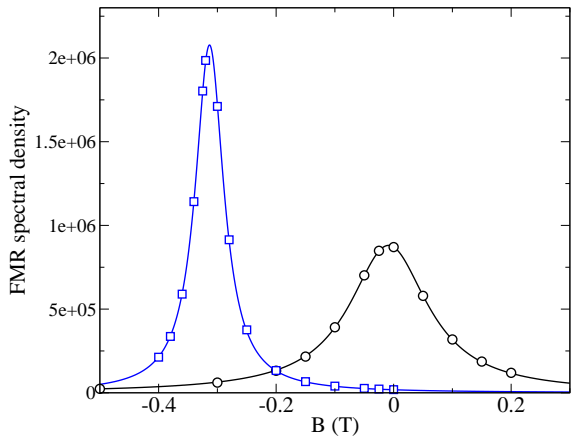}
\caption{Spectral densitty of the FMR mode for the 2Co/2Pt system
plotted as a function of the Zeeman field $B$ at a fixed pumping frequencies:
$\nu_p=24$~GHz (squares) and $\nu_p=54$~GHz (circles). The solid curves are Lorentzian fits to the calculated points.} 
\label{chi_of_B}
\end{figure}

Figure~\ref{alpha_x_eta_x_NCo} shows the dependence of the Gilbert damping constant
$\alpha_G$ on the imaginary part $\eta$ for Co films of various thicknesses. Clearly
$\alpha_G$ approaches finite values as $\eta\rightarrow 0$. 
Cobalt has a  small spin-orbit coupling constant. We
would like to investigate the effect of 
increasing the strength of the SOC on the damping rate. Instead of artificially increasing
$\xi$ in Co we  consider a more realistic setting where a double layer of Co
is attached to a non-magnetic substrate with high SOC parameter, such as Pt. This system
has a particularly interesting feature: the magnetization easy axis is perpendicular to the plane. 
However, we found that, for the LCAO parameters we employed, the magnetization in-plane is 
also a stable configuration,
with a small magnetocrystalline anisotropy.
The damping rate, however, is much larger in the 2Co/2Pt
system than in the unsupported Co films. This is a nice example of how the anisotropy energy 
is strongly influenced by the system's symmetry, but the damping rate is relatively insensitive to it, 
depending strongly on
the intensity of the spin-orbit coupling. It is also an extremely convenient situation to test
an assumption very frequently found in the literature on Gilbert damping, although sometimes 
not explicitly stated: that the FMR linewidth $\Delta\Omega$ is linearly dependent on the resonance 
frequency $\Omega_0$ and that $\Delta\Omega\rightarrow 0$ as
$\Omega_0\rightarrow 0$. This is not an unreasonable hypothesis, 
considering the weak static fields commonly used in FMR experiments 
and the smallness of the spin-orbit coupling constant, compared to
other energy scales of a ferromagnet. Our calculations 
for the Co films confirm that this relationship is approximately held. 
In this case, the Gilbert constant $\alpha_G$ may be extracted from 
the FMR spectrum by simply fitting it to a Lorentzian and is practically field-independent.
However, our results for 2Co/2Pt indicate that the FMR linewidth
is finite as $\Omega_0\rightarrow 0$, leading to a significantly 
frequency-dependent $\alpha_G$, as shown in Fig.~\ref{DW_x_W0_2Co2Pt}.
In order to illustrate how the determination of a damping parameter is affected by
the finite value of $\Delta\Omega$ as $\Omega_0\rightarrow 0$ we extracted
the linewidths from the calculated spectra for the 2Co/2Pt system by fitting Lorentzians
to our calculated spectral densities. The results are shown
in Fig.~\ref{DW_x_W0_2Co2Pt}. One of its most important consequences
is that, if one wishes to define a value of $\alpha_G$ for the system above,
it must be defined as a function of the Zeeman field, as is illustrated in
Fig.~\ref{DW_x_W0_2Co2Pt}. In principle this poses a problem for
the procedure usually employed to determine FMR spectra experimentally, since there
the free variable is the Zeeman field, not the frequency of the exciting field.
In Fig.~\ref{chi_of_B} we illustrate this issue by plotting the FMR spectral
density as a function of the Zeeman field for two fixed pumping frequencies, 24~GHz and 54~GHz. 
The curves have nice Lorentzian shapes, but the values for the Gilbert damping 
parameter $\alpha_G$ extracted from these curves depend on the pumping frequency
($\alpha_G=0.034$ for $\Omega_0=0.10$~meV and $\alpha_G=0.042$ for $\Omega_0=0.22$~meV).
Also, they do not correspond to any of the values shown in 
Fig.~\ref{DW_x_W0_2Co2Pt}b, although the Zeeman field values that determine
the linewidth in Fig.~\ref{chi_of_B} lie within the range of Zeeman field values 
showed in Fig.~\ref{DW_x_W0_2Co2Pt}b. Thus, if $\alpha_G$ is defined as  $\Delta\Omega/\Omega_0$,
its value for a given sample depends on wether the FMR spectrum is obtained in a fixed frequency
or fixed Zeeman field set ups.
Our results also imply that the existing expressions for the damping constant $\alpha_G$ 
are not valid in general, specially for very clean systems with large spin-orbit coupling materials.
The conventional approaches express $\alpha_G$
as the ratio $\Delta\Omega/\Omega_0$ in the $\Omega_0\rightarrow 0$ limit. As we have just shown,
this limit does not exist in general, since $\Delta\Omega$ approaches a finite value as
$\Omega_0\rightarrow 0$.

In experimental papers~\cite{Woltersdorf_Heinrich_PRL_2001,Heinrich_PRL_2014} 
the FMR linewidth is assumed to have a zero-frequency offset, just as we described. This is usually
attributed to extrinsic broadening mechanisms, such as two-magnon scattering~\cite{Arias_Mills_two_magnon}, 
due to the combination between inhomogeneities in the magnetic films and dipolar interactions. This is
certainly the case in systems with small SOC, such as Fe films deposited on GaAs or Au~\cite{Woltersdorf_Heinrich_PRL_2001}.
However, we have shown that there can be zero-frequency offset of intrinsic origin if the SOC is large. 
The effect of this intrinsic offset  should be easily separated from that of the two-magnon scattering mechanism, 
since the latter is not active when the magnetization is perpendicular to the plane of the film~\cite{Arias_Mills_two_magnon}.

\begin{figure}
\includegraphics[width=\columnwidth]{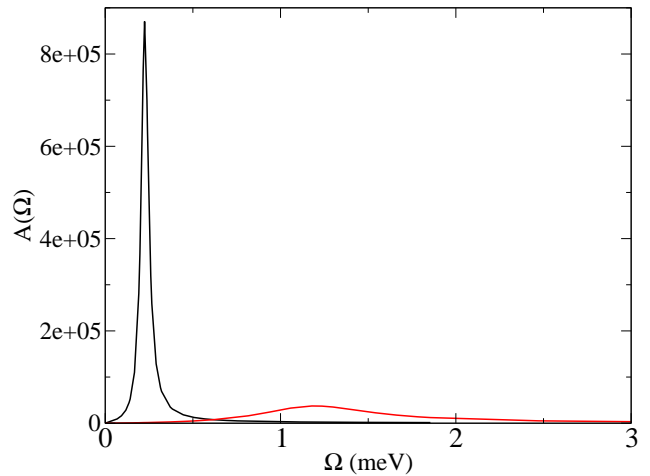}
\caption{a) Spectral densities of the FMR mode for the 2Co/2Pt system
with Stoner enhancement in Pt turned on (black line)  and off (red line).}
\label{enh_x_nonenh}
\end{figure}

We would like to remark  that Stoner enhancement in Pt plays a very important 
role in the determination of the damping rate. We had shown previously~\cite{spinpumpingPd} 
that, in the absence of spin-orbit coupling, Stoner enhancement had a very mild effect on the 
damping rate in the Co/Pd(001) system. In the presence of SOC, however, the effect can be very large 
indeed. Both magnetocrystalline anisotropy and damping rate are significantly different in the 
enhanced and non-enhanced cases, as shown in Fig.~\ref{enh_x_nonenh}. The Gilbert parameter is
also very different in the two cases: $\alpha_G^\mathrm{enh}=0.11$, whereas 
$\alpha_G^\mathrm{non-enh}=0.33$. Thus, proper treatment of Stoner enhancement 
in substrates like Pd an Pt is essential for the correct determination of spin 
relaxation features.
%


We presented a microscopic approach to the calculation of the Gilbert 
damping parameter $\alpha_G$ for ultrathin metallic magnetic films,
illustrated by results for Co films and Co/Pt bilayers. 
Our approach is based on the evaluation of the 
dynamic transverse susceptibility in the presence of spin-orbit coupling,
taking into account realistic electronic structures and the coupling
between transverse spin, longitudinal spin and charge excitations. It
predicts finite values of $\alpha_G$ in the limit of perfectly crystalline films,
a regime where methods based on the torque correlation
formula find a diverging Gilbert damping parameter. 
We showed that the coupling between transverse, longitudinal and charge excitations, 
due to spin-orbit coupling, is of fundamental importance for
the correct determination of FMR spectra in metallic systems. 
We have also shown that the damping rate extracted from the FMR spectrum
for fixed pumping frequency differs considerably from that extracted from the FMR 
spectrum for fixed Zeeman field. In this case the Gilbert damping parameter $\alpha_G$
becomes frequency dependent, in contrast to what is assumed in the standard Landau-Lifshitz-Gilbert
phenomenology.
Moreover, we have numerical indications that
the Gilbert parameter is not well defined in the limit
of vanishing resonance frequency, a fact that is very relevant to 
calculational schemes based on the adiabatic approximation. 
Incidentally, Stoner enhancement in materials like Pt and Pd also plays
an important role in the determination of FMR frequencies and damping rates.
These results may lead to important modifications
of the interpretation of damping ``constants'', either calculated or
inferred from experimental results, for systems where
spin-orbit coupling is strong. 
We believe these issues may be crucial for the correct
description of relaxation in very clean systems of nanoscopic dimensions,
specially in the presence of relatively weak magnetocrystalline anisotropy.

The authors acknowledge partial financial support from CNPq and FAPERJ. We are
grateful to Professor Caio Lewenkopf for a critical reading of the manuscript and to 
Dr. Mariana Odashima for enlightening discussions. RBM acknowledges fruitful discussions with Prof. D. M. Edwards and A.Umerski. 

\bibliography{gilbert}

\begin{thebibliography}{15}
\expandafter\ifx\csname natexlab\endcsname\relax\def\natexlab#1{#1}\fi
\expandafter\ifx\csname bibnamefont\endcsname\relax
  \def\bibnamefont#1{#1}\fi
\expandafter\ifx\csname bibfnamefont\endcsname\relax
  \def\bibfnamefont#1{#1}\fi
\expandafter\ifx\csname citenamefont\endcsname\relax
  \def\citenamefont#1{#1}\fi
\expandafter\ifx\csname url\endcsname\relax
  \def\url#1{\texttt{#1}}\fi
\expandafter\ifx\csname urlprefix\endcsname\relax\def\urlprefix{URL }\fi
\providecommand{\bibinfo}[2]{#2}
\providecommand{\eprint}[2][]{\url{#2}}

\bibitem[{\citenamefont{Gilbert}(2004)}]{gilbert_ieee}
\bibinfo{author}{\bibfnamefont{T.}~\bibnamefont{Gilbert}},
  \bibinfo{journal}{Magnetics, IEEE Transactions on}
  \textbf{\bibinfo{volume}{40}}, \bibinfo{pages}{3443} (\bibinfo{year}{2004}),
  ISSN \bibinfo{issn}{0018-9464}.

\bibitem[{\citenamefont{Tserkovnyak et~al.}(2005)\citenamefont{Tserkovnyak,
  Brataas, Bauer, and Halperin}}]{Bauer_RevModPhys}
\bibinfo{author}{\bibfnamefont{Y.}~\bibnamefont{Tserkovnyak}},
  \bibinfo{author}{\bibfnamefont{A.}~\bibnamefont{Brataas}},
  \bibinfo{author}{\bibfnamefont{G.}~\bibnamefont{Bauer}}, \bibnamefont{and}
  \bibinfo{author}{\bibfnamefont{B.}~\bibnamefont{Halperin}},
  \bibinfo{journal}{Rev. Mod. Phys.} \textbf{\bibinfo{volume}{77}},
  \bibinfo{pages}{1375} (\bibinfo{year}{2005}),
  \urlprefix\url{http://link.aps.org/doi/10.1103/RevModPhys.77.1375}.

\bibitem[{\citenamefont{Kambersk\'y}(2007)}]{Kambersky2007}
\bibinfo{author}{\bibfnamefont{V.}~\bibnamefont{Kambersk\'y}},
  \bibinfo{journal}{Phys. Rev. B} \textbf{\bibinfo{volume}{76}},
  \bibinfo{pages}{134416} (\bibinfo{year}{2007}),
  \urlprefix\url{http://link.aps.org/doi/10.1103/PhysRevB.76.134416}.

\bibitem[{\citenamefont{Garate and
  MacDonald}(2009{\natexlab{a}})}]{GarateMacdonald_2009I}
\bibinfo{author}{\bibfnamefont{I.}~\bibnamefont{Garate}} \bibnamefont{and}
  \bibinfo{author}{\bibfnamefont{A.}~\bibnamefont{MacDonald}},
  \bibinfo{journal}{Phys. Rev. B} \textbf{\bibinfo{volume}{79}},
  \bibinfo{pages}{064403} (\bibinfo{year}{2009}{\natexlab{a}}),
  \urlprefix\url{http://link.aps.org/doi/10.1103/PhysRevB.79.064403}.

\bibitem[{\citenamefont{Garate and
  MacDonald}(2009{\natexlab{b}})}]{GarateMacdonald_2009II}
\bibinfo{author}{\bibfnamefont{I.}~\bibnamefont{Garate}} \bibnamefont{and}
  \bibinfo{author}{\bibfnamefont{A.}~\bibnamefont{MacDonald}},
  \bibinfo{journal}{Phys. Rev. B} \textbf{\bibinfo{volume}{79}},
  \bibinfo{pages}{064404} (\bibinfo{year}{2009}{\natexlab{b}}),
  \urlprefix\url{http://link.aps.org/doi/10.1103/PhysRevB.79.064404}.

\bibitem[{\citenamefont{Starikov et~al.}(2010)\citenamefont{Starikov, Kelly,
  Brataas, Tserkovnyak, and Bauer}}]{Bauer_Tserkovnyak_Kelly_2010}
\bibinfo{author}{\bibfnamefont{A.}~\bibnamefont{Starikov}},
  \bibinfo{author}{\bibfnamefont{P.}~\bibnamefont{Kelly}},
  \bibinfo{author}{\bibfnamefont{A.}~\bibnamefont{Brataas}},
  \bibinfo{author}{\bibfnamefont{Y.}~\bibnamefont{Tserkovnyak}},
  \bibnamefont{and} \bibinfo{author}{\bibfnamefont{G.}~\bibnamefont{Bauer}},
  \bibinfo{journal}{Phys. Rev. Lett.} \textbf{\bibinfo{volume}{105}},
  \bibinfo{pages}{236601} (\bibinfo{year}{2010}),
  \urlprefix\url{http://link.aps.org/doi/10.1103/PhysRevLett.105.236601}.

\bibitem[{\citenamefont{Ebert et~al.}(2011)\citenamefont{Ebert, Mankovsky,
  K\"odderitzsch, and Kelly}}]{Kelly_2011}
\bibinfo{author}{\bibfnamefont{H.}~\bibnamefont{Ebert}},
  \bibinfo{author}{\bibfnamefont{S.}~\bibnamefont{Mankovsky}},
  \bibinfo{author}{\bibfnamefont{D.}~\bibnamefont{K\"odderitzsch}},
  \bibnamefont{and} \bibinfo{author}{\bibfnamefont{P.~J.} \bibnamefont{Kelly}},
  \bibinfo{journal}{Phys. Rev. Lett.} \textbf{\bibinfo{volume}{107}},
  \bibinfo{pages}{066603} (\bibinfo{year}{2011}),
  \urlprefix\url{http://link.aps.org/doi/10.1103/PhysRevLett.107.066603}.

\bibitem[{\citenamefont{Barati et~al.}(2014)\citenamefont{Barati, Cinal,
  Edwards, and Umerski}}]{Edwards_2014}
\bibinfo{author}{\bibfnamefont{E.}~\bibnamefont{Barati}},
  \bibinfo{author}{\bibfnamefont{M.}~\bibnamefont{Cinal}},
  \bibinfo{author}{\bibfnamefont{D.~M.} \bibnamefont{Edwards}},
  \bibnamefont{and} \bibinfo{author}{\bibfnamefont{A.}~\bibnamefont{Umerski}},
  \bibinfo{journal}{Phys. Rev. B} \textbf{\bibinfo{volume}{90}},
  \bibinfo{pages}{014420} (\bibinfo{year}{2014}),
  \urlprefix\url{http://link.aps.org/doi/10.1103/PhysRevB.90.014420}.

\bibitem[{\citenamefont{Muniz and Mills}(2003)}]{MunizMills2003}
\bibinfo{author}{\bibfnamefont{R.~B.} \bibnamefont{Muniz}} \bibnamefont{and}
  \bibinfo{author}{\bibfnamefont{D.~L.} \bibnamefont{Mills}},
  \bibinfo{journal}{Phys. Rev. B} \textbf{\bibinfo{volume}{68}},
  \bibinfo{pages}{224414} (\bibinfo{year}{2003}),
  \urlprefix\url{http://link.aps.org/doi/10.1103/PhysRevB.68.224414}.

\bibitem[{\citenamefont{Costa et~al.}(2010)\citenamefont{Costa, Muniz, Lounis,
  Klautau, and Mills}}]{Costa_SOC_2010}
\bibinfo{author}{\bibfnamefont{A.~T.} \bibnamefont{Costa}},
  \bibinfo{author}{\bibfnamefont{R.~B.} \bibnamefont{Muniz}},
  \bibinfo{author}{\bibfnamefont{S.}~\bibnamefont{Lounis}},
  \bibinfo{author}{\bibfnamefont{A.~B.} \bibnamefont{Klautau}},
  \bibnamefont{and} \bibinfo{author}{\bibfnamefont{D.~L.} \bibnamefont{Mills}},
  \bibinfo{journal}{Phys. Rev. B} \textbf{\bibinfo{volume}{82}},
  \bibinfo{pages}{014428} (\bibinfo{year}{2010}),
  \urlprefix\url{http://link.aps.org/doi/10.1103/PhysRevB.82.014428}.

\bibitem[{\citenamefont{Liu et~al.}(2014)\citenamefont{Liu, Yuan, Wesselink,
  Starikov, and Kelly}}]{Kelly_PRL2014}
\bibinfo{author}{\bibfnamefont{Y.}~\bibnamefont{Liu}},
  \bibinfo{author}{\bibfnamefont{Z.}~\bibnamefont{Yuan}},
  \bibinfo{author}{\bibfnamefont{R.~J.~H.} \bibnamefont{Wesselink}},
  \bibinfo{author}{\bibfnamefont{A.~A.} \bibnamefont{Starikov}},
  \bibnamefont{and} \bibinfo{author}{\bibfnamefont{P.~J.} \bibnamefont{Kelly}},
  \bibinfo{journal}{Phys. Rev. Lett.} \textbf{\bibinfo{volume}{113}},
  \bibinfo{pages}{207202} (\bibinfo{year}{2014}),
  \urlprefix\url{http://link.aps.org/doi/10.1103/PhysRevLett.113.207202}.

\bibitem[{\citenamefont{Urban et~al.}(2001)\citenamefont{Urban, Woltersdorf,
  and Heinrich}}]{Woltersdorf_Heinrich_PRL_2001}
\bibinfo{author}{\bibfnamefont{R.}~\bibnamefont{Urban}},
  \bibinfo{author}{\bibfnamefont{G.}~\bibnamefont{Woltersdorf}},
  \bibnamefont{and} \bibinfo{author}{\bibfnamefont{B.}~\bibnamefont{Heinrich}},
  \bibinfo{journal}{Phys. Rev. Lett.} \textbf{\bibinfo{volume}{87}},
  \bibinfo{pages}{217204} (\bibinfo{year}{2001}),
  \urlprefix\url{http://link.aps.org/doi/10.1103/PhysRevLett.87.217204}.

\bibitem[{\citenamefont{Montoya et~al.}(2014)\citenamefont{Montoya, Heinrich,
  and Girt}}]{Heinrich_PRL_2014}
\bibinfo{author}{\bibfnamefont{E.}~\bibnamefont{Montoya}},
  \bibinfo{author}{\bibfnamefont{B.}~\bibnamefont{Heinrich}}, \bibnamefont{and}
  \bibinfo{author}{\bibfnamefont{E.}~\bibnamefont{Girt}},
  \bibinfo{journal}{Phys. Rev. Lett.} \textbf{\bibinfo{volume}{113}},
  \bibinfo{pages}{136601} (\bibinfo{year}{2014}),
  \urlprefix\url{http://link.aps.org/doi/10.1103/PhysRevLett.113.136601}.

\bibitem[{\citenamefont{Arias and Mills}(1999)}]{Arias_Mills_two_magnon}
\bibinfo{author}{\bibfnamefont{R.}~\bibnamefont{Arias}} \bibnamefont{and}
  \bibinfo{author}{\bibfnamefont{D.}~\bibnamefont{Mills}},
  \bibinfo{journal}{Phys. Rev. B} \textbf{\bibinfo{volume}{60}},
  \bibinfo{pages}{7395} (\bibinfo{year}{1999}),
  \urlprefix\url{http://link.aps.org/doi/10.1103/PhysRevB.60.7395}.

\bibitem[{\citenamefont{Santos et~al.}(2013)\citenamefont{Santos, Venezuela,
  Muniz, and Costa}}]{spinpumpingPd}
\bibinfo{author}{\bibfnamefont{D.~L.~R.} \bibnamefont{Santos}},
  \bibinfo{author}{\bibfnamefont{P.}~\bibnamefont{Venezuela}},
  \bibinfo{author}{\bibfnamefont{R.~B.} \bibnamefont{Muniz}}, \bibnamefont{and}
  \bibinfo{author}{\bibfnamefont{A.~T.} \bibnamefont{Costa}},
  \bibinfo{journal}{Phys. Rev. B} \textbf{\bibinfo{volume}{88}},
  \bibinfo{pages}{054423} (\bibinfo{year}{2013}),
  \urlprefix\url{http://link.aps.org/doi/10.1103/PhysRevB.88.054423}.

\end{thebibliography}

\end{document}